\documentclass[twocolumn,prd,showpacs,preprintnumbers,amsmath,amssymb,floatfix]{revtex4}

\usepackage{graphicx}

\newcommand{\beq}{\begin{equation}}
\newcommand{\eeq}{\end{equation}}
\newcommand{\beqa}{\begin{eqnarray}}
\newcommand{\eeqa}{\end{eqnarray}}

\newcommand{\sign}{\mathop{\mathrm{sign}}}
\newcommand{\Nt}{\tilde N}

\newcommand{\mpl}{m_{\rm pl}}

\newcommand{\const}{\mathop{\mathrm{const}}}
\newcommand{\sh}{_{\scriptstyle H}}
\newcommand{\sv}{_{\scriptstyle V}}

\newcommand{\figref}[1]{Fig.~\ref{#1}}
\newcommand{\Figref}[1]{Figure~\ref{#1}}

\begin{document}

\title{On the accuracy of slow-roll inflation given
current observational constraints} 

\author{Alexey Makarov}
\email{amakarov@princeton.edu}
\affiliation{Physics Department, Princeton University, Princeton, NJ 08544}

\date{\today}

\begin{abstract}
We investigate the accuracy of slow-roll inflation in light of current
observational constraints, which do not allow for a large deviation
from scale invariance.  We investigate the applicability of the first
and second order slow-roll approximations for inflationary models,
including those with large running of the scalar spectral index.  We
compare the full numerical solutions with those given by the first and
second order slow-roll formulae.  We find that even first order
slow-roll is generally accurate; the largest deviations arise in
models with large running where the error in the power spectrum can be
at the level of 1-2\%. Most of this error comes from inaccuracy in the
calculation of the slope and not of the running or higher order terms.
Second order slow-roll does not improve the accuracy over first order.
We also argue that in the basis $\epsilon_0=1/H$,
$\epsilon_{n+1}={d\ln|\epsilon_n|}/{dN}$, introduced by Schwarz et
al. (2001), slow-roll does not require all of the parameters to be
small. For example, even a divergent $\epsilon_3$ leads to finite
solutions which are accurately described by a slow-roll approximation.
Finally, we argue that power spectrum parametrization recently
introduced by Abazaijan, Kadota and Stewart does not work for models
where spectral index changes from red to blue, while the usual Taylor
expansion remains a good approximation.

\end{abstract}


\pacs{98.80.Cq}
\clearpage

\maketitle

\newlength\pictwidth
\setlength\pictwidth{3.4in}
\newlength\widepictwidth
\setlength\widepictwidth{7.0in}

\section{Introduction}

Inflation is a theory which postulates that a rapid expansion of the universe
occurred right after the Big Bang 
\cite{1981PhRvD..23..347G,1981MNRAS.195..467S,1982PhRvL..48.1220A,1982PhLB..108..389L}. 
Most inflationary models can be represented by an effective single
field model with effective potential $V$. The inflaton
with mass $m$ rolls down the potential until the kinetic energy
of the inflaton is greater than half of its potential energy. 
At this point the inflationary expansion of the universe
stops and the next phase of reheating occurs.
During the inflationary expansion, the initial quantum fluctuations
exponentially increase and become classical 
\cite{1979JETPL..30..682S,1981JETPL..33..532M,1982PhRvL..49.1110G,%
1983PhRvD..28..679B,1982PhLB..115..295H,1982PhLB..117..175S}.
These classical fluctuations also seed the subsequent 
growth of large scale structure.
There is a well defined procedure which allows us to find the spectrum of the 
fluctuations given the inflationary potential. 
Because exact solutions are numerically intensive several appoximations
have been developed. 
The most common approximation is the slow-roll approximation.
Recently the so-called uniform approximation 
was suggested  \citep{2004PhRvD..70h3507H,2005PhRvD..71d3518H}.
Reference \cite{2005PhRvD..71d3517C} developed improved WKB-type 
approximation.

If the kinetic energy of the inflaton is much smaller than
its potential energy, we say that the inflaton is slowly rolling 
down its potential. In this slow-roll approximation
we can obtain analytical formulae for the produced power spectrum
in the form of a Taylor series expansion in a set of slow-roll parameters.
The coefficients in the Taylor expansion of the logarithm 
of the power spectrum in $\ln k$ effectively 
define the slope $n_s-1$, running $\alpha_s$ and
higher derivatives.
We usually derive the slow-roll formulae through the time delay formalism
or Bessel function approximation. Therefore there are some
implied conditions on the accuracy of the slow-roll approximation
depending upon the slow-roll parameters.
References \cite{1997PhLB..414...18W,2002PhRvD..66b3515L}
found that there are areas in the slow-roll parameter space
where the accuracy of slow-roll approximation is questionable.
This usually requires a large deviation of $n_s$ from 1. 
However it contradicts the latest observations 
\citep{2003ApJS..148..175S,2003ApJS..148..195V,2005PhRvD..71j3515S}.

Recently there has been a lot of renewed interest in models with large running
of the scalar index \cite{2003ApJS..148..213P,2003PhRvD..68b3508K,2003PhRvD..68f3501C}. 
It is not clear whether slow-roll approximation is accurate in this
area of parameter space, as in some expansions one of the slow 
roll parameters becomes large and the expansion is no longer 
well controlled \cite{2002PhRvD..66b3515L}. Another issue is the question of 
where to stop the expansion. 
Although it is often assumed that the running is $O\left((n_s-1)^2\right)$,
reference \cite{2002PhRvD..65j1301D} found that there are cases where it can
be as large as $n_s-1$.
In this case one should also consider the effect of
including the running of the running of $n_s$, i.e. the second derivative
of $n_s$ over $\ln k$. 
These are the issues addressed in this paper.
We begin with a short review of the basic physics of inflation and
the algorithm of numerical solutions
to the inflationary equations, with more details given in appendix.
We continue by comparing 
the numerical solutions to those given by slow-roll approximations 
and finally we present our conclusions. 

In this paper we use a standard convention for reduced Planck mass
$\mpl=G_N^{-1/2}$.

\section{Inflationary basics}
In the ``Hamilton-Jacobi'' formulation, the evolution 
of the Hubble parameter $H(\phi)$ during inflation
with potential $V(\phi)$ is given by (e.g. see \cite{2000cils.conf.....L})
\beq
\left[H'(\phi)\right]^2-\frac{12\pi}{\mpl^2}H^2(\phi)=-\frac{32\pi^2}{\mpl^4}V(\phi).
\label{eq:jacobi}
\eeq

The number of e-folds $N$ since some initial time is related 
to the value of the scalar field $\phi$ by
\beq
\frac{dN}{d\phi}=-\frac{4\pi}{\mpl^2}\frac{H(\phi)}{H'(\phi)}.
\eeq
We will consider the situation when the value of the scalar field is growing in time, $d\phi/dt>0$.
Then by our convention $dN/dt$ is also positive, $dN/dt>0$.

In the literature, different sets of slow-roll parameters are used.
Reference \cite{1992PhLB..291..391L} introduce potential slow-roll parameters which are 
constructed on the basis of the derivatives of the inflationary 
potential $V(\phi)$.
Authors of \cite{1994PhRvD..50.7222L} define Hubble slow-roll parameters
through the derivatives of the Hubble parameter $H(\phi)$ 
with respect to the field $\phi$ during inflation 
\beqa
\epsilon\sh(\phi)&=&\dfrac{\mpl^2}{4\pi}\left(\frac{H'(\phi)}{H(\phi)}\right)^2,\\
\eta\sh(\phi)&=&\dfrac{\mpl^2}{4\pi}\frac{H''(\phi)}{H(\phi)},\\
^n\xi\sh(\phi)&=&\left(\dfrac{\mpl^2}{4\pi}\right)^n\left.\frac{(H')^{n-1}H^{(n+1)}}{H^n}\right..
\label{eq:HSRP}
\eeqa
In this parameterization when the inequality 
$\epsilon\sh(\phi)<1$ fails, the inflation immediately stops.
Sometimes $^2\xi\sh$ is also denoted $\xi\sh$ or $\xi\sh^2$ though
it can take negative values. In this paper we will use $\xi\sh\equiv\left.^2\xi\sh\right.$.

Reference \cite{2001PhLB..517..243S} introduces another basis of ``horizon-flow'' slow-roll parameters
through the logarithmic derivative of the Hubble distance $\epsilon_0=d\sh=1/H(N)$ with respect to the
number of e-folds $N$ to the end of inflation
\beq
\epsilon_{n+1}=\dfrac{d\ln|\epsilon_n|}{dN}.
\label{eq:flow-horizon}
\eeq
The connection between any two of these sets can be found
in e.g. \cite{2001PhLB..517..243S}.
Thus the first three horizon-flow slow-roll parameters are connected to the first 
three Hubble slow-roll parameters as  
\citep{1994PhRvD..50.7222L,1997RvMP...69..373L,2002PhRvD..66b3515L}
\beqa
\epsilon_1&=&\epsilon\sh,\\
\epsilon_2&=&2\epsilon\sh-2\eta\sh,\\
\epsilon_2\epsilon_3&=&4\epsilon\sh^2-6\epsilon\sh\eta\sh+2\xi\sh.
\eeqa
There is an analytical connection between
Hubble slow-roll parameters and potential slow-roll 
parameters \citep{1994PhRvD..50.7222L}. 

References \cite{2000PhRvD..62j3520M,1993PhLB..302..171S,2001PhLB..510....1S,2004PhLB..603...95W} 
use differently defined 
sets of slow-roll parameters, but they still can be converted to the ones we have described here
(e.g. see \citep{2001PhLB..517..243S}).

Thus any inflationary model can be completely described by the evolution of one 
of the sets of the parameters. 

The condition for the inflation to occur is $\epsilon_1=\epsilon\sh<1$ or
$\epsilon\sv\lesssim1$ since $\epsilon\sh=\epsilon\sv$ to first order.

To find the power spectrum of the perturbations produced by a single field inflation,
one can follow the prescription of Grivell and Liddle \citep{1996PhRvD..54.7191G}.
One solves the 
equation \citep{1985JETPL..41..493M,1988ZhETF..94....1M,1993PhLB..302..171S}
\beq
\frac{d^2u_k}{d\tau^2}+\left(k^2-\frac1z\frac{d^2z}{d\tau^2}\right)u_k=0
\label{eq:modeeq}
\eeq
for each mode with wavenumber $k$ and initial condition
$
u_k(\tau)\to\frac1{\sqrt{2k}}e^{-ik\tau}
$
as $\tau\to-\infty$. Then the spectrum of curvature perturbations is given by
\beq
\mathcal P_{\mathcal R}(k)=\frac{k^3}{2\pi^2}\left|\frac{u_k}z\right|^2.
\eeq
The quantity $z$ in equation~\eqref{eq:modeeq}
is defined as $z=a\dot\phi/H$ for scalar modes and $z=a$ for tensor modes.
Then for scalar modes \citep{1996PhRvD..54.7191G}
\begin{multline}
\frac1z\frac{d^2z}{d\tau^2}=2a^2H^2\left[1+\epsilon\sh-\frac32\eta\sh\right.\\
\left.+\epsilon\sh^2-2\epsilon\sh\eta\sh+\frac12\eta\sh^2+\frac12\xi\sh\right].
\label{eq:d2zdt2}
\end{multline}

One can parametrize the power spectrum of the scalar and tensor modes of
the fluctuations amplified by the inflation as 
\beq
\ln\frac{\mathcal P(k)}{\mathcal P_0}=
(n-1)\ln\frac k{k_*}
+\frac{\alpha}2\ln^2\frac k{k_*}
+\frac{\beta}6\ln^3\frac k{k_*}+\dots
\label{eq:parameterization}
\eeq
around some conventional pivot point $k_*$.
Leach et al. \citep{2002PhRvD..66b3515L} give expressions for
the scalar spectral index $n_s$, 
the running of the scalar spectral index $\alpha_s$,
the tensor spectral index $n_t$ and 
the running of the tensor spectral index $\alpha_t$
in terms of the horizon-flow parameters. Here we reproduce 
their second order formulae for $n_s-1$ and $\alpha_s$
\beqa
n_s-1&=&-2\epsilon_1-\epsilon_2-2\epsilon_1^2\notag\\
&&-(2C+3)\epsilon_1\epsilon_2
-C\epsilon_2\epsilon_3,\label{eq:nsm1}\\
\alpha_s&=&-2\epsilon_1\epsilon_2-\epsilon_2\epsilon_3,
\label{eq:running}
\eeqa
where $C=\gamma_{\mathrm E}+\ln2-2\approx-0.7296$.

Reference \cite{2002PhRvD..66b3515L} also analyzes
the  accuracy of the approximation~\eqref{eq:parameterization}
for parameterizing the inflationary power spectrum
of fluctuations with $\beta=0$ 
for different values of the parameters $r$, $n_s$ and $\alpha_s$.

In this paper we will also compare the second order formulae (\ref{eq:nsm1}, \ref{eq:running})
to the first order formulae given by
\beqa
n_s-1&=&-2\epsilon_1-\epsilon_2,\label{eq:nsm1_first}\\
\alpha_s&=&-2\epsilon_1\epsilon_2-\epsilon_2\epsilon_3.
\label{eq:running_first}
\eeqa
The expression for $\alpha_s$ is the same as in the second order formula
because the expression for $\alpha_s$ is derived using only first order expression for $n_s$.
Thus the main difference between first and second order formulae comes from the extra terms in the expression for $n_s-1$.


Current observational constraints on $r$, $n_s$ and $\alpha_s$ 
are given by \cite{2004PhRvD..69j3501T,2005PhRvD..71j3515S,2004PhRvD..69l3003S}.
At 95\% confidence level, the tensor to scalar ratio is $R<0.50$, 
which implies that the first horizon-flow parameter 
$\epsilon_1$ is much smaller than one. 
Current constraints on the scalar spectral index give us
$n_s=0.98\pm0.02$, which 
in turn means that the second horizon-flow
slow-roll parameter is much smaller than one. 

Present data does not require the presence
of running in the primordial 
power spectrum \citep{2004MNRAS.351L..49L}, but running as large
as $\pm 0.03$ is still allowed at 3-$\sigma$ \citep{2005PhRvD..71j3515S}. 
Regular inflationary models usually predict $|\alpha_s| \sim (n_s-1)^2$
and so the running is of the order of $10^{-3}$, 
as is the case for the minimally-coupled $V(\phi)=\lambda\phi^4$ model 
with 60 e-folds remaining.

\begin{figure}
\includegraphics[width=\pictwidth]{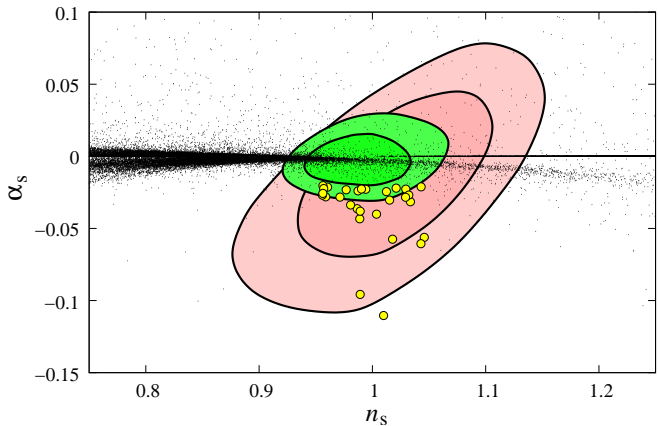}
\caption{\label{fig:fig1}
The current constraints 
(68\% and 95\% confidence level contours) in 
the $n_s$--$\alpha_s$ plane from WMAP+SDSSgal 
(bigger, red) and WMAP+SDSSlya (smaller, green) data 
\citep{2005PhRvD..71j3515S,2004PhRvD..69l3003S,2004PhRvD..69j3501T,
2003ApJS..148..195V}.
The constrained region clearly allows the value of the spectral scalar index 
$n_s$ to be around $1$ and the running $\alpha_s$ of the scalar spectral index
to be significantly non-zero for either combination of the experiments. }
\end{figure}

But it is possible that $|\alpha_s|\gg(n_s-1)^2$ and $\alpha_s<0$,
which means that the main part in the running 
of the spectral index~\eqref{eq:running}
is determined not by the first term $-2\epsilon_1\epsilon_2$, 
but by the second term $-\epsilon_2\epsilon_3$.
It happens when $|\epsilon_3|\gg |\epsilon_1|$, and therefore there might be 
a situation when $|\epsilon_3|>1$.

To summarize, if $n_s\approx1$ and $\alpha_s$ is a small negative number, 
at some scale we might have $\epsilon_1\ll1$,
$\epsilon_2\ll1$ and $|\epsilon_3|>1$. 
Leach et al. \citep{2002PhRvD..66b3515L} define inflation satisfying slow-roll
under the condition $|\epsilon_n|\ll1$, for all $n>0$.
In our case $\epsilon_3>1$, so the question arises as to whether
slow-roll in this case is accurate or whether the 
approximation breaks down and one must also include 
terms with higher powers in $\epsilon_3$.
Does it mean that the inflation
is not slow-roll and one must use full numerical solutions instead? 
And does it mean that one must also include the running of the running? 
These are the main questions we address in this paper. 
To address them 
we have developed the numerical code described in appendix A.

How natural is it for inflation with a given number from 50 to 70 e-folds remaining 
to produce a power spectrum with a changing tilt? 
In the absence of theoretical guidance on the inflationary space
we cannot address this question simply. 
Authors of \cite{2003ApJS..148..213P} have produced about 200,000 simulations
of the inflationary flow equations for more or less ``random'' potentials, and 
calculated the observable parameters ($n_s$, $\alpha_s$, $r$, $n_t$, $\alpha_t$) 
of the resulting power spectra about 40 to 70 e-folds before the end of the inflation for each potential. 
About 80,000 of them fall into the area plotted on
\figref{fig:fig1}. Only the fifteen marked with larger yellow circles 
give a significant change in the tilt from red to blue, i.e. $n_s \sim 1$, $\alpha_s<-0.02$. 

Choosing the Hubble parameter to be represented 
by a Taylor expansion in $\phi$ with 
uniformly distributed coefficients, as done in \cite{2003ApJS..148..213P},
does not 
necessarily correspond to the real inflationary priors \citep{2003PhRvD..68j3504L}. 
We do not address this issue here; instead we want to simply
stress
that possibility of constructing a potential with a large running
in the scalar power spectrum 40-70 e-folds 
before the end of the inflation exists.

\section{Quadratic potential}
\begin{figure}
\includegraphics[width=\pictwidth]{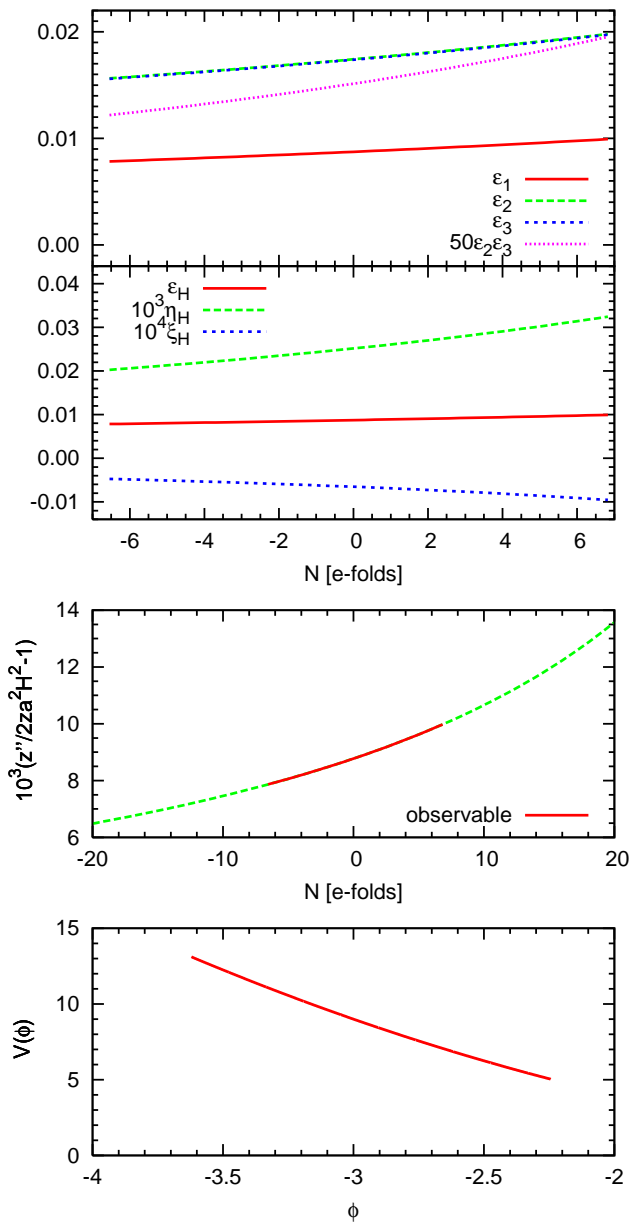}
\caption{\label{fig2_phi2}
Panels from the bottom to the top: 
1.~Potential $V=m^2\phi^2/2$ in the range of $\phi$'s where the inflation occurs
around 50 e-folds before the end of inflation. 
The scale in $V$ is not COBE-normalized on this plot.
2.~The dependence of $z''/z$ on the number of the e-folds during the inflation.
Number of e-folds $N=0$ corresponds to our arbitrarily chosen pivot scale of $k=0.05$/Mpc.
3.~Plots of Hubble slow-roll parameters $\epsilon\sh$, $\eta\sh$, $\xi\sh$.
The value of $\epsilon\sh$ is non-negligible whereas $\eta\sh$ and $\xi\sh$
are essentially zeros. 
4.~Plots of horizon-flow slow-roll parameters $\epsilon_1$, $\epsilon_2$,
$\epsilon_3$ and the product of $\epsilon_2\epsilon_3$ 
for the same inflationary model. Since in this case
$\epsilon\sh\gg\eta\sh,\xi\sh$, the value of $\epsilon_1$ is essentially
$\epsilon_1\approx\epsilon\sh$ and
$\epsilon_2\approx\epsilon_3\approx2\epsilon\sh$. Nothing unexpected is going on
here for this model.
}
\end{figure}

As a test of our code, in this section we investigate
how well the slow-roll formulae work in
the slow-roll regime for one of the usual potentials that do not 
predict large running. As an example we will consider a simple quadratic
potential, $V=m^2\phi^2/2$, which is a classic example of chaotic
inflation. The second panel from the bottom in \figref{fig2_phi2} effectively shows the dependence 
of $z''/z$ on the number of the e-folds for inflation with such a
potential. The behavior is monotonic and very smooth, which is due to the
smoothness of the derivatives of the potential. Since $z''/z$ scales as $2a^2H^2$, 
we plot the quantity 
\beq
\frac1{2a^2H^2}\frac{z''}z-1
\eeq
instead (compare to equation~\eqref{eq:d2zdt2}).

The top two panels show the dependence of $\epsilon\sh$, $\eta\sh$ and
$\xi\sh$ on the number of e-folds. The only significantly non-zero term is
$\epsilon\sh$, which gradually grows to $1$ at the end of 
inflation. The values of $\eta\sh$ and $\xi\sh$ are typically smaller by roughly $10^3$ and
$10^4$ respectively.

\begin{figure}
\includegraphics[width=\pictwidth]{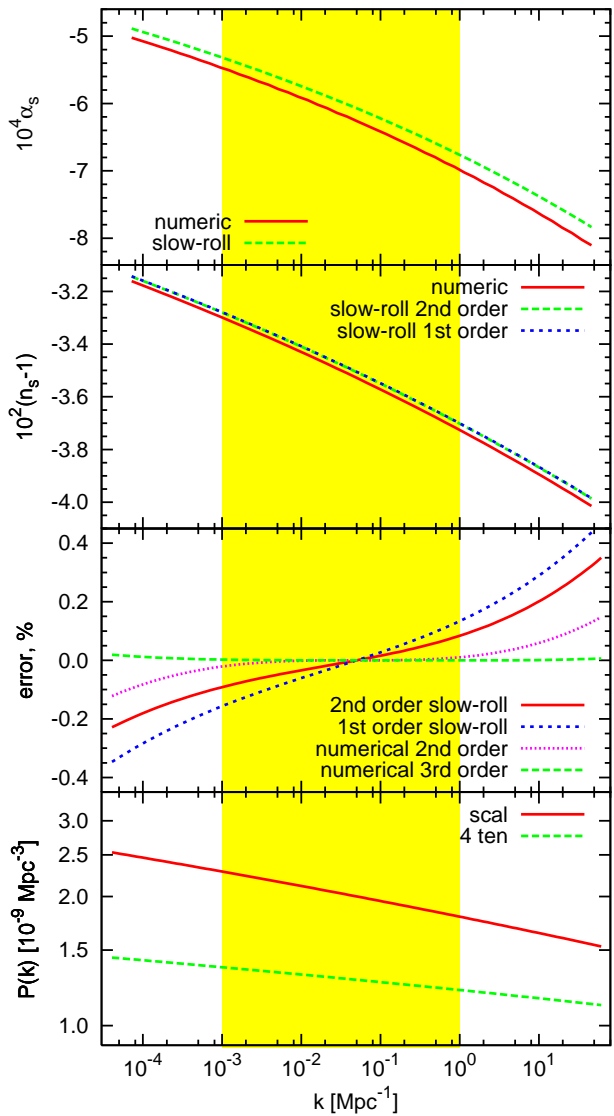}
\caption{\label{fig3_phi2}
Panels from the bottom to the top: 
1.~Power spectrum produced by inflation with potential $V=m^2\phi^2/2$. 
2.~Errors made by approximating the true scalar power spectrum by \eqref{eq:parameterization} with
calculating $n_s$ and $\alpha_s$ from first order slow-roll formulae (\ref{eq:nsm1_first}, \ref{eq:running_first}),
second order slow-roll formulae (\ref{eq:nsm1}, \ref{eq:running}) and calculating them through numerical
derivatives. Numerical third order takes into account third logarithmic derivative of the power spectrum $\beta_s$
in parameterization \eqref{eq:parameterization}. 
3.~The evolution of $n_s-1$ is calculated numerically and by the first and second order slow-roll formulae.
4.~The evolution of $\alpha_s$ is calculated numerically and by the slow-roll formulae 
(first and second order slow-roll are the same for $\alpha_s$).
The error plot clearly shows that the main error comes from the imprecision of $n_s-1$, whereas 
the approximation for the $\alpha_s$ works well enough. One can also see that in the case of this potential 
both first and second order slow-roll formulae for $n_s-1$ overestimate the
real value of $n_s-1$.
}
\end{figure}

\Figref{fig3_phi2} shows the primordial power spectrum produced by the quadratic potential.
The second panel from the bottom describes the error produced by the slow-roll approximations.
The first order approximation gives less than 0.2\% error in the observed range of wavenumbers $k$.
The second order approximation works slightly better; the error is just above 0.1\%. 
Both of these numbers
are likely to be good enough for the upcoming experiments. 
This is because the accuracy at large scales is limited by the 
finite number of modes, while at small scales it is limited by the nonlinear 
evolution. So, while the overall amplitude could in principle be determined 
to an accuracy of 0.1\% when CMB and lensing information is combined, 
it is unlikely that such a precision will be achieved separately at two 
widely separated length scales. 

Taking a more careful look at the error plot, one sees that
the error curve in the observed area is basically a straight line,
meaning that the main source of error is not the imprecise value of $\alpha_s$
but the error in $n_s$. Let us estimate now
how precisely we need to know $n_s$ to get an error of, say 0.2\%, in the observed range.
The imprecision $\delta n_s$ in $n_s$ will give us the uncertainty
\beq
\delta n_s \frac12\ln \frac{k_{\max}}{k_{\min}}=0.002.
\eeq
Taking the observed range of $k$'s to be from $10^{-3}$~Mpc$^{-1}$ to $1$~Mpc$^{-1}$, we find that
one needs to find $n_s$ with the precision of $\delta n_s=6\cdot10^{-4}$.

The same allowed uncertainty $\delta\alpha_s$ in $\alpha_s$ is estimated from
\beq
\frac12\delta\alpha_s\left(\frac12\ln \frac{k_{\max}}{k_{\min}}\right)^2=0.002.
\eeq
Therefore $\delta\alpha_s=3\cdot10^{-4}$ is the error 
which we can make in determining $\alpha_s$ in order to get 
an error in the power spectrum of 0.2\% at the edges of the observed range of $k$'s.

The two top panels of \figref{fig3_phi2} compare the numerically found
dependence of $n_s$ and $\alpha_s$ on $k$ to the one found from the slow-roll approximation
with the first order $\alpha_s$ and either the first or second order for $n_s$.
One should
compare the discrepancies between these to the values
of $\delta n_s$ and $\delta\alpha_s$. The characteristic value
of $n_s$ is .964 and the discrepancy between the exact value and the 
one found from the slow-roll approximation is comparable to $\delta n_s$. 
Running $\alpha_s$ takes values around $-6.5\cdot10^{-4}$. The discrepancy
between the exact and the slow-roll values is very small in
comparison to $\delta\alpha_s$. One can also notice that
in this case $|\alpha_s|\approx2\delta\alpha_s$. 
Thus,
even if we assigned $\alpha_s=0$, we would not get 
a significant error in the approximation of the primordial
power spectrum of the scalar perturbations.

To summarize this section, for standard inflationary potentials, 
the slow-roll approximation suffices even at first order when compared to 
the expected accuracy of existing and future experiments. 
The second order approximation, while improving the accuracy, is not really necessary. 
The main error of slow-roll when considered in contrast to the numerical solutions
is the inaccuracy in the slope $n_s$; inaccuracies in higher 
order expansion terms, such as the running, are less important and can even 
be ignored. 

\section{Potential with a bump in the second derivative}
We want to construct a potential which will give us
a strong running and crossing of the point $n_s=1$ 
in the observable power spectrum.
We want to have $n_s>1$ at earlier times in inflation, while
at later times we want to have $n_s<1$. To get
the desired result, one can take two different potentials
producing such features and smoothly connect them. 

One can rewrite slow-roll formulae (\ref{eq:nsm1},\ref{eq:running}) through
the potential slow-roll parameters as
\beqa
n_s-1&=&-6\epsilon\sv+2\eta\sv,\\
\alpha_s&=&16\epsilon\sv\eta\sv-24\epsilon\sv^2-2\xi\sv.
\eeqa
Now let us just choose our potential to be 
\beq
V(\phi)=1-0.01\phi-1.20\phi^2
\label{eq:pot1}
\eeq
for all $\phi>0$. 
This choice provides about 50 e-folds of inflation after $\phi=0$.
Since the local properties of the power spectrum are mostly 
determined by the local ``history'' of the slow-roll parameters 
at the moment of the horizon crossing, we
can get a red tilt of the scalar power spectrum 
$n_s\approx0.80$ in the area where
the ``history'' before point $\phi=0$ is not very important.
To get an approximately symmetric shape of the power spectrum we choose
$V(\phi)$ to be
\beq
V(\phi)=1-0.01\phi+1.20\phi^2
\label{eq:pot2}
\eeq
for all $\phi<0$. In this case for wave modes which cross the horizon
far before the moment when the scalar field
takes the value of $\phi=0$, the spectral index
of the primordial power spectrum has a blue tilt $n_s\approx1.20$.
Thus between these two regions the
spectral index changes from 1.20 to 0.80.
We can unite formulae~\eqref{eq:pot1} and~\eqref{eq:pot2} into
\beq
V(\phi)=1-0.01\phi-1.20\phi^2\sign\phi.
\label{eq:singpotential}
\eeq

\begin{figure}
\includegraphics[width=\pictwidth]{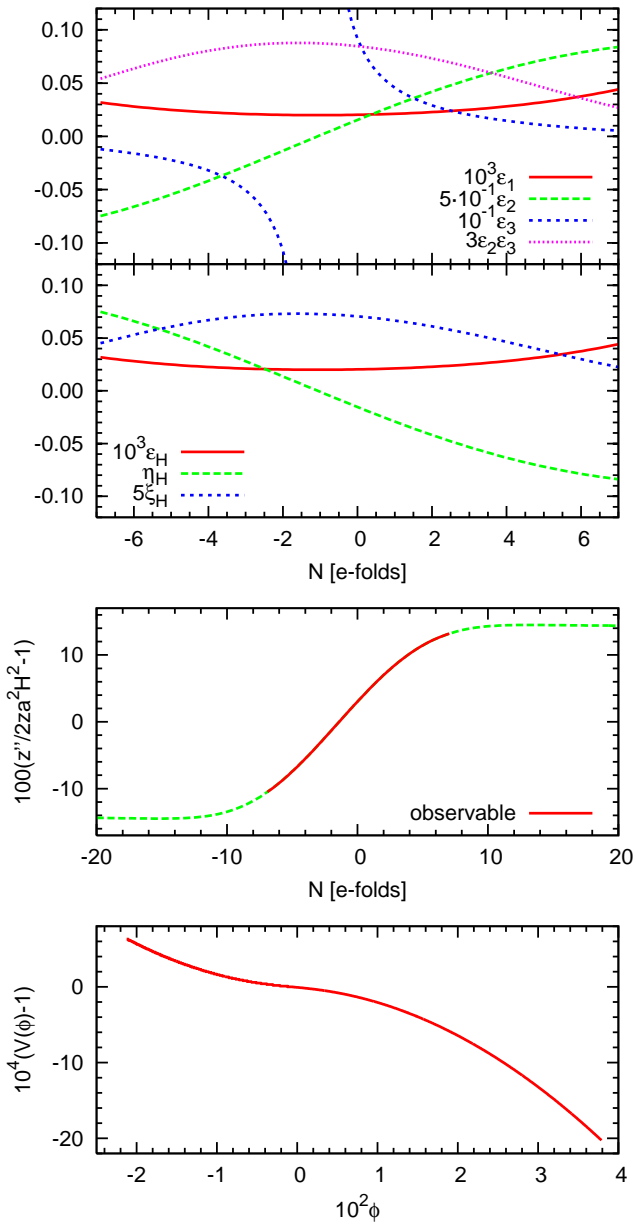}
\caption{\label{fig2}
Panels from the bottom to the top: 
1.~Potential~\eqref{eq:smoothedpotential}.
2.~Effectively this plot shows the dependence of $z''/z$ 
on the number of the e-folds during the inflation.
3.~Plots of Hubble slow-roll parameters $\epsilon\sh$, $\eta\sh$, $\xi\sh$.
Though the potential has a singular behavior, all the flow parameters are smooth.
4.~Plots of horizon-flow slow-roll parameters $\epsilon_1$, $\epsilon_2$,
$\epsilon_3$ and the product of $\epsilon_2\epsilon_3$ 
for the same inflationary model. While everything 
is fine with $\epsilon_1$, $\epsilon_2$ and $\epsilon_2\epsilon_3$, 
the value of $\epsilon_3$ indeed flips over infinity.}
\end{figure}

\begin{figure}
\includegraphics[width=\pictwidth]{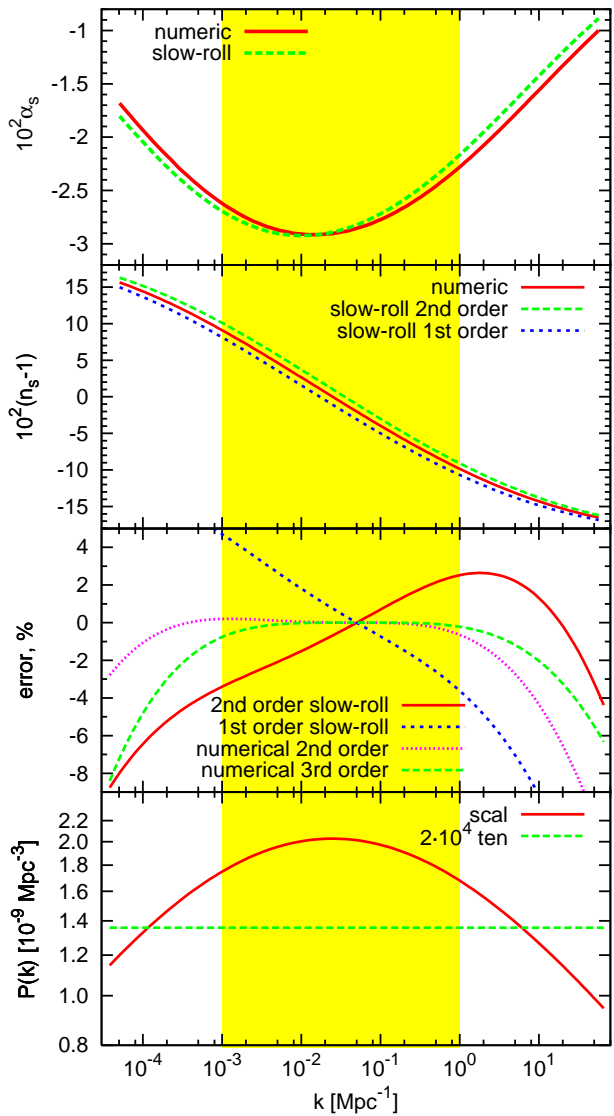}
\caption{\label{fig3}
Panels from the bottom to the top: 
1.~Power spectrum produced by~\eqref{eq:smoothedpotential}.
2.~Errors made by approximating the true scalar power spectrum 
by \eqref{eq:parameterization} with
calculating $n_s$ and $\alpha_s$ from first order 
slow-roll formulae (\ref{eq:nsm1_first}, \ref{eq:running_first}),
second order slow-roll formulae (\ref{eq:nsm1}, \ref{eq:running}) 
and calculating them through numerical
differentiation. Numerical third order takes into 
account third logarithmic derivative of the power spectrum $\beta_s$
in parameterization \eqref{eq:parameterization}.
3.~The evolution of $n_s-1$ is calculated numerically 
and by the first and second order slow-roll formulae.
4.~The evolution of $\alpha_s$ is calculated numerically 
and by the slow-roll formulae (first and second order slow-roll are the same for $\alpha_s$).
The error plot clearly shows that the main error comes from the imprecision of $n_s-1$, whereas 
the approximation for the $\alpha_s$ works well enough. 
We see that in the case of this potential, 
first order slow-roll formula for $n_s-1$ underestimates the real value, while the second
order formula overestimates it.
}
\end{figure}

This potential has continuous first and second derivatives,
but has a bump in its third derivative.
This makes $\dfrac1z\dfrac{d^2z}{d\tau^2}$ 
in~\eqref{eq:d2zdt2} discontinuous around 
$\phi=0$. According to \cite{1992PZETF..55..477S}
this produces oscillations in the power spectrum,
which we can indeed see for the potential~\eqref{eq:singpotential}.
To avoid the oscillations we smooth out
the $\sign\phi$ function, changing 
it to $\dfrac2\pi\arctan(200\phi)$. 
In this case~\eqref{eq:singpotential} changes to
\beq
V(\phi)=1-0.01\phi-1.20\phi^2\dfrac2\pi\arctan(200\phi),
\label{eq:smoothedpotential}
\eeq
which is shown on \figref{fig2}.
We have chosen 200 as the coefficient in front of $\phi$ 
in the $\arctan$ function so that the produced power spectrum
has a nice shape as in \figref{fig3}.

The two top panels of \figref{fig2} show the behavior of
the slow-roll parameters $\epsilon\sh$, $\eta\sh$, $\xi\sh$
and $\epsilon_1$, $\epsilon_2$, $\epsilon_3$, 
$\epsilon_2\epsilon_3$ correspondingly.
While nothing unexpected happens to the behavior 
of the conventional Hubble slow-roll parameters $\epsilon\sh$, $\eta\sh$ and $\xi\sh$,
there appears to be a singularity for the horizon-flow parameter $\epsilon_3$.
However, notice that the product $\epsilon_2\epsilon_3$
behaves smoothly and remains small 
due to the fact that the parameter $\epsilon_2$
is changing its sign and therefore crossing through zero.
Thus the parameterization of equation~\eqref{eq:flow-horizon} introduces
a singularity which is not physically present in the model.

\Figref{fig3} shows the power spectrum produced by 
the model of inflation with the potential~\eqref{eq:singpotential}.
The second panel from the bottom shows the errors made by different
approximations. We again observe a similar picture for the slow-roll
formulae. The main source of error for either 
the first or second order approximations comes not from the value of $\alpha_s$
but from the error in the value of $n_s$. From the second panel from the
top, we can estimate that the discrepancy is of the order of 0.01 for 
\beq
\delta n_s=n_s^{\mathrm{exact}}-n_s^{\mathrm{approx}}
\eeq
which gives an error of 
\beq
\delta n_s \frac12\ln \frac{k_{\max}}{k_{\min}}\approx 4\%
\eeq
in the produced power spectrum at the edges of the observed range.
Both the first and second order slow-roll approximations for $n_s$
work somewhat unsatisfactory. 
The first order slow roll underestimates $n_s$ and
the second order overestimates it by about the same amount.

On the other hand, if our goal is to focus on running alone 
regardless of the slope and 
use just that property to deduce something about the 
potential, then the slow-roll does very well, since 
the differences between the slow-roll and numerical value of running 
are very small even at the lowest order in slow-roll. 
Extra terms in the expansion~\eqref{eq:parameterization} further 
improve the accuracy. Adding running of the running 
improves the accuracy over the observed range from 
1\% to 0.2\%. 

In summary, for potentials that lead to large running, 
slow-roll does not estimate the slope $n_s$ very accurately
at either first or second order, while the accuracy of the running $\alpha_s$
suffices for the existing and future experiments. If we observe
over a wide range of scales then it is useful to add the cubic term. 
Second order slow-roll does not seem to improve the accuracy. 

\section{Flow Equations Simulations}
\label{sec:flow}
Kinney \cite{2002PhRvD..66h3508K} introduced a formalism based on the so-called
flow equations, further discussed in 
\cite{2003PhRvD..68j3504L}.
The basic idea is that if one
fixes the Hubble slow-roll parameters~\eqref{eq:HSRP} at some point
in time for $\epsilon\sh$, $\eta\sh$ and $^\ell\xi\sh$ up to $\ell=M$ 
and  assumes that all the other Hubble slow-roll
parameters are small enough that one can neglect them in one's
calculations (i.e. $^\ell\xi\sh=0$ for all $\ell\ge M+1$) then,
without any other assumptions about inflation being slow-roll, one
can find the Hubble slow-roll parameters at any other
moment of time using the following hierarchy of linear 
ordinary differential equations:
\beqa
\frac{d\epsilon}{dN}&=&-2\epsilon(\eta-\epsilon),\label{eq:flow1}\\
\frac{d\eta}{dN}&=&\epsilon\eta-~^2\xi,\\
\frac{d~^\ell\xi}{dN}&=&[\ell\epsilon-(\ell-1)\eta]~^\ell\xi -~ ^{\ell+1}\xi
\label{eq:flowlast}
\eeqa
for all $\ell=2\dots M$ assuming $^{M+1}\xi=0$.

Usually when we set up an inflationary problem, we choose
a potential $V(\phi)$ and then reconstruct the 
form of the Hubble parameter during inflation
using the main nonperturbed Hamilton-Jacobi inflationary equation~\eqref{eq:jacobi},
which gives us an attractor solution $H(\phi)$ which in the inflationary class
of problems almost does not depend on the initial condition.

By following the method prescribed by \cite{2002PhRvD..66h3508K}
one avoids solving 
the main attractor inflationary equation~\eqref{eq:jacobi},
as pointed out by \cite{2003PhRvD..68j3504L}. Indeed,
the assumption $^\ell\xi\sh=0$ for all $\ell\ge M+1$
requires that $H^{(\ell)}(\phi)=0$ for all $\ell\ge M+2$.
Consequently, $H(\phi)$ is a polynomial of order $M+1$:
\beq
H(\phi)=H_0(1+A_1\phi+A_2\phi^2+A_3\phi^3+\dots+A_{M+1}\phi^{M+1}).
\eeq
In this case the function $H(\phi)$ is an attractor solution of equation~\eqref{eq:jacobi}
with a potential in the form
\begin{multline}
V(\phi)=-\frac{\mpl^4}{32\pi^2}\left([H'(\phi)]^2-\dfrac{12\pi}{\mpl^2}H^2(\phi)\right)\\
=-\frac{\mpl^4}{32\pi^2}H_0^2\Biggl[\left(A_1+\dots+(M+1)A_{M+1}\phi^{M}\right)^2 \\
-\dfrac{12\pi}{\mpl^2}\left(1+A_1\phi+\dots+A_{M+1}\phi^{M+1}\right)^2 \Biggr].
\end{multline}

Thus the only differential equation one needs to solve in order to match up
the number of e-folds and the value of the scalar field $\phi$ 
is 
\beq
\frac{dN}{d\phi}=\frac{2\sqrt{\pi}}{\mpl}\frac1{\sqrt{\epsilon(\phi)}}.
\label{eq:dNdphi}
\eeq
Here again $\epsilon(\phi)$ is defined as in the equation~\eqref{eq:HSRP}:
\begin{multline}
\epsilon(\phi)=\frac{\mpl^2}{4\pi}\left(\frac{H'(\phi)}{H(\phi)}\right)^2\\
=\frac{\mpl^2}{4\pi}\left(\frac{A_1+2A_2\phi+\dots+(M+1)A_{M+1}\phi^{M}}
{1+A_1\phi+A_2\phi^2+\dots+A_{M+1}\phi^{M+1}}\right)^2.
\end{multline}

We do not have to numerically solve the hierarchy of $M$ differential 
flow equations. Instead we have analytical expressions for $V(\phi)$ and $H(\phi)$.

The late attractor $\epsilon= {}^\ell\xi=0$ and $\eta=\const$, found by \cite{2002PhRvD..66h3508K},
corresponds to the situation where the inflation proceeds to the value
of the scalar field $\phi$, which is a solution of the equation $\epsilon=0$
\beq
A_1+2A_2\phi+\dots+(M+1)A_{M+1}\phi^{M}=0.
\eeq
At this point if $A_2\neq0$, then $\eta=\const\neq0$ due to the definition of $\eta$, which
does not involve $\epsilon$ at all. All the other $^\ell\xi=0$ since any of them
is a product of the first derivative of the Hubble parameter (which is zero) 
with some higher order derivatives.

Peiris et al. \cite{2003ApJS..148..213P} made $M=9$-th order flow equation simulations; about 
$40,000$ are shown as black dots on \figref{fig:fig1}. Fifty points
fall into the range $|n_s-1|<0.05$ and $\alpha_s<-0.02$; these are shown
in yellow. Among these point we have chosen 13 which fall into the narrow
interval $|n_s-1|<0.02$, and we have reconstructed
the corresponding inflationary potentials for the inflationary models which
give such significant running, together with $n_s$ extremely close to $1$.

\begin{figure}
\includegraphics[width=\pictwidth]{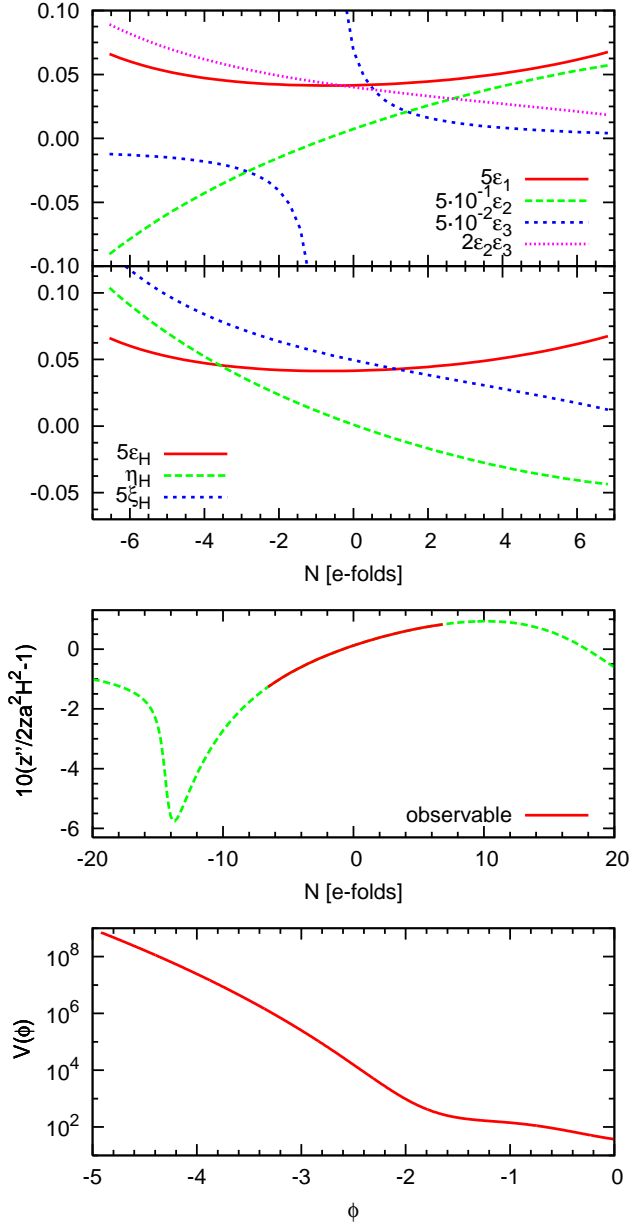}
\caption{\label{fig4}
Panels from the bottom to the top: 
1.~Potential reconstructed from one of the flow equations simulations.
2.~Effectively this plot shows the dependence of $z''/z$
on the number of the e-folds during the inflation for this potential.
3.~Plots of Hubble slow-roll parameters $\epsilon\sh$, $\eta\sh$, $\xi\sh$.
All of them are smooth.
4.~Plots of horizon-flow slow-roll parameters $\epsilon_1$, $\epsilon_2$,
$\epsilon_3$ and the product of $\epsilon_2\epsilon_3$ 
for the same inflationary model. While everything 
is fine with $\epsilon_1$, $\epsilon_2$ and $\epsilon_2\epsilon_3$, 
the value of $\epsilon_3$ again flips over infinity.}
\end{figure}

\begin{figure}
\includegraphics[width=\pictwidth]{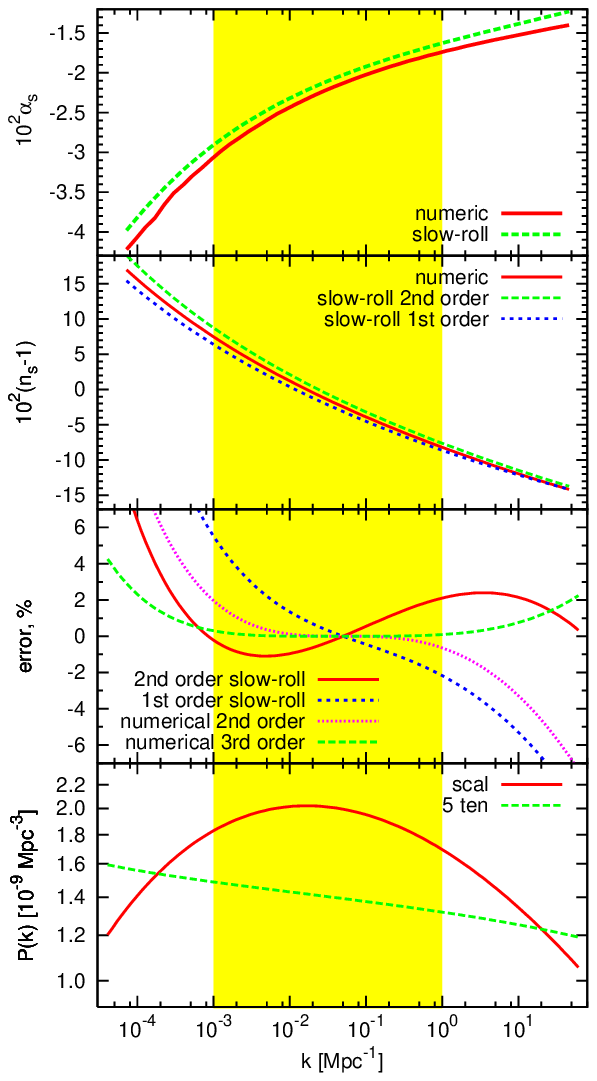}
\caption{\label{fig5}
Panels from the bottom to the top: 
1.~Power spectrum produced by potential on \figref{fig4}.
2.~Errors made by approximating the true scalar power spectrum by \eqref{eq:parameterization} with
calculating $n_s$ and $\alpha_s$ from first order slow-roll formulae (\ref{eq:nsm1_first}, \ref{eq:running_first}),
second order slow-roll formulae (\ref{eq:nsm1}, \ref{eq:running}) and calculating them through numerical
differentiation. Numerical third order takes into account third logarithmic derivative of the power spectrum $\beta_s$
in parameterization \eqref{eq:parameterization}.
3.~The evolution of $n_s-1$ is calculated numerically and by the first and second order slow-roll formulae.
4.~The evolution of $\alpha_s$ is calculated numerically and  by the slow-roll formulae (first and second order slow-roll are the same for $\alpha_s$).
The error plot clearly shows that the main error comes from the imprecision of $n_s-1$, whereas 
the approximation for the $\alpha_s$ works well enough. One can also see that in the case of this potential 
first order slow-roll formula for $n_s-1$ underestimates the real value, while the second
order formula overestimates it.}
\end{figure}

The bottom panel in \figref{fig4} shows a potential from such a model
with an unusually high value of the running $\alpha_s$. We notice that
there is a small dip
in the potential. Some of the potentials with high $\alpha_s$ from the
simulations had unrealistically high values of the tensor
to the scalar ratio, but all of them had quite similar shapes.
The second from the bottom panel of \figref{fig4} shows the characteristic behavior
of the function $z''/z$ which influences the scalar power spectrum
as we have seen earlier.
The top two panels show the dependence of the slow-roll parameters on the number of e-folds.
As in the other case with large running, we find a singularity
for the horizon-flow slow-roll parameter $\epsilon_3$, while 
the product $\epsilon_2\epsilon_3$ behaves smoothly and $\epsilon_2$ crosses
zero.

The bottom panel in \figref{fig5} shows the power spectrum of scalar
and tensor perturbations produced by inflation with the potential under consideration.
The second from the bottom panel shows the error produced by every one 
of the approximations for the power spectrum. We again see that both
first and second order slow-roll formulae do not give a satisfactory
result for $n_s$. One of them again overestimates $n_s$; 
the other underestimates it. The error for either of the approximations
is about 2--4\%. 

The error introduced by the approximate formula for the running $\alpha_s$
is a bit smaller than the one for $n_s$, but it is
somewhat larger compared to the quadratic potential we
considered in the previous section.

Chen et al. \citep{2004CQGra..21.3223C} perform a similar analysis
of slow-roll approximation. Using the flow-equations technique, they
found discrepancy of larger than $0.01$ for $n_s$
between second and third order slow-roll approximations for some of the models. Based on this fact 
they conclude that third order slow-roll is better. For the model
we considered in this section
we have found that the third order slow-roll does not improve the results of the
second order approximation. In our calculations both formulas give identical results leading 
to approximately the same order of error as the first order approximation.

\section{Is Truncated Taylor Expansion good?}
Recently Abazajian, Kadota and Stewart \citep{2005astro.ph..7224A} have argued that if 
\beq
|\alpha_s \ln(k/k_*)| \gtrsim |n_s-1|,
\label{eq:trtrTa}
\eeq 
then the traditional truncated Taylor series parameterization is inconsistent,
and hence it can lead to incorrect parameter estimations. One can notice that Taylor
expansions $P(x)=\sum a_ix^i$ of functions $x^2$ or $\cos x$ around $x=0$ also violates
the condition $a_1\gtrsim a_2x$, but no one argues that these expansions are not valid.
Abazajian et al. propose to use the parameterization
\beq
\ln\mathcal P(k)=\ln\mathcal P_0+\frac{(n_s-1)^2}{\alpha_s}
\left[\left(\frac{k}{k_*}\right)^{\frac{\alpha_s}{n_s-1}}-1\right]
\label{eq:AKS}
\eeq
instead.

There is one significant disadvantage of this approach. In particular,
using this parameterization to describe $\mathcal P$ as a function of
$k$, one is able to describe only a growing or decreasing function,
which can be easily seen from the form of the function. The models we
study in this paper produce scalar power spectra which are not purely
growing or decreasing (e.g. see \figref{fig5}).

\begin{figure}
\includegraphics[width=\pictwidth]{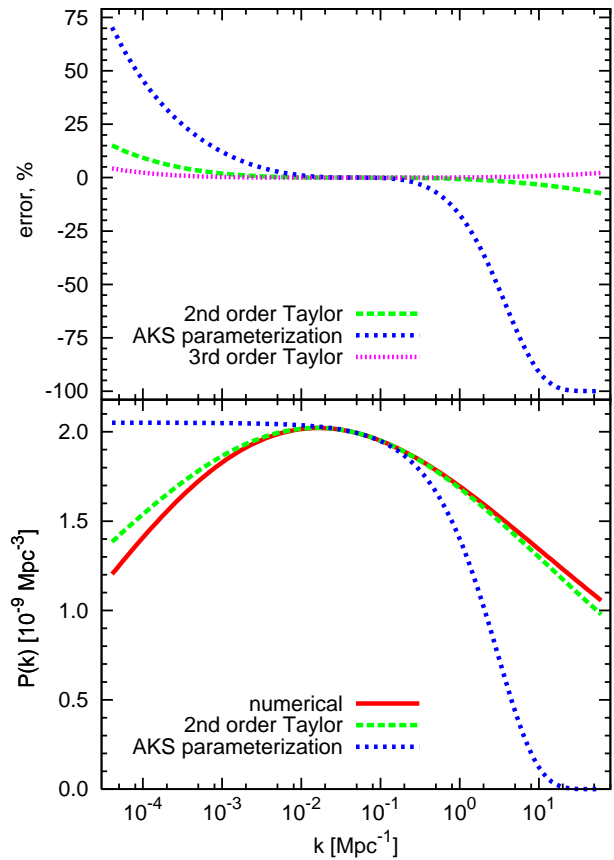}
\caption{\label{fig:aks}
The bottom panel shows the numerically calculated power spectrum of
the potential we used in section~\ref{sec:flow} and it's
approximations by the second order Taylor expansion and by Abazajian,
Kadota and Stewart (AKS) parameterization~\eqref{eq:AKS}.  The top
panel shows the error produced by each of the parameterizations.  We
have also added the error produced by the third order Taylor expansion
here.  AKS parameterization acceptably describes the true power
spectrum in a very narrow range of $k$'s around the pivot point
$k=0.05$~Mpc$^{-1}$.  It does not improve the truncated Taylor
expansion over the wider range of $k$'s.
}
\end{figure}

In the previous section we have considered the potential which
produces power spectrum satisfying equation~\eqref{eq:trtrTa}. On
\figref{fig:aks} we compare the traditional truncated to second and
third order Taylor expansion and the parameterization~\eqref{eq:AKS}. We
find that the parameterization~\eqref{eq:AKS} gives a significantly
larger error than, e.g. the second order Taylor expansion.

Thus we found that in this particular case though equation~\eqref{eq:trtrTa} holds,
truncated Taylor expansion is a good approximation and the AKS approach does not 
improve it. There might be models for which equation~\eqref{eq:AKS} works better than
Taylor expansion, but it is definitely not an improvement for a general case
and should be used with caution, if at all.

\section{Conclusions}

In this paper we have explored the accuracy of the slow-roll approximation
given the observational constraints
on the primordial scalar and tensor power spectra. The current constraints
can be roughly described by the tensor to scalar ratio $r<1$, 
small deviation from the scale invariance of the scalar power spectrum, 
$|n_s-1|<0.05$ and small but possibly nontrivial 
running,
$|\alpha_s|<0.03$. These constraints allow for the particular case where $n_s \sim 1$
and $|\alpha_s|>0.01$, which has previously been argued to not satisfy the 
slow-roll condition. We have computed exact numerical solutions for the 
considered potentials and
compared them to those obtained from the first and second order slow-roll
approximations.

We have found that for the potentials explored here,
 there is no substantial difference
when using first or second order slow-roll formulae
for the power spectrum index $n_s$. Both of them either work well
in the case of small running or have a comparable error 
in the case of non-negligible running.
Adding extra (cubic in $\ln k$) terms in the approximation for
the scalar power spectrum extends the accuracy to a larger range of 
scales, but this accuracy is most likely not necessary for existing and near
future experiments. 
If the values of $n_s$ and $\alpha_s$ are known
with the precision $\delta n_s=6\cdot10^{-4}$ and $\delta\alpha_s=3\cdot10^{-4}$,
then the scalar power spectrum will have
an error of about 0.2\% at the edge of the observable range of wavenumbers $k$'s.

The horizon-flow basis $\epsilon_{n+1}={d\ln|\epsilon_n|}/{dN}$
introduces an artificial singularity for inflationary 
models with negative running and the value of the spectral index crossing 1.
Such a divergence in one of the horizon-flow parameters does not indicate 
that the slow-roll approximation has been badly broken. 
We find that the slow-roll is still accurate at the 1-2\% level and most of 
the error comes from inaccuracies in the evaluation of the slope itself, 
and not the running. 
Thus the first order slow-roll
approximation
is sufficiently accurate for the current observations. 
Only if the running turns out to be large, while the slope remains close
to scale-invariant, are exact numerical calculations 
required to achieve sub-percent accuracy. In the appendix we present a short guideline
on performing such calculations.  One can request the code directly from the author.

\begin{acknowledgments}
The author is thankful to Hiranya Peiris for making the results of her 
Monte Carlo simulations available. 
AM also thanks Sergei Bashinsky, Chris Beasley, Latham Boyle, 
Steven Gratton, Patricia Li, Uro\v s Seljak and Alexei Starobinskii for useful discussions and comments.
\end{acknowledgments}

\bibliography{apjmnemonic,cosmo,cosmo_preprints,additional}   

\appendix*
\section{Inflationary Equations}
In this appendix we describe the technical details of the code 
we ran to get the results presented in the main part of the paper.
The code is given a potential $V(\phi)$ and some point $\phi_0$
which lies in the observable range of wave-modes and, say,
corresponds to the moment when wavelengths with $k=0.05$~Mpc$^{-1}$
exit the horizon. We want to find the power spectrum produced
by inflation with the potential $V(\phi)$. For this purpose
we first have to go backwards in time about 50 e-folds and then start
the inflation there. This guarantees that the inflationary
dynamics are not affected by the choice of the initial condition and
we indeed have the attractor solution.

Now we evolve the universe from our ``beginning of inflation''
to the end of inflation, the moment which is determined by the violation
of the inequality $\ddot a>0$. This part is described below in the
``non-perturbed inflationary equations'' section. 
Usually we require 50 to 70 e-folds between $\phi_0$ and the end of inflation.

After we already have the complete background history of 
the evolution of the universe
during the inflationary stage of the expansion,
we can start working out the evolution of the
perturbations during inflation, as discussed in the second part of the appendix.

\subsection{Non-perturbed inflationary equations}
The unperturbed dynamics of inflation are described
by the equation of motion of the scalar field $\phi$ with potential
$V(\phi)$ in the expanding universe with 
the Hubble parameter $H\equiv\dot a/a$
\beq
\ddot\phi+3H\dot\phi+V'(\phi)=0
\label{eq:infl1}
\eeq
and the Friedman equation with only the scalar field component 
present in the universe
\beq
H^2=\frac{8\pi}{3\mpl^2}\left[V(\phi)+\frac12\dot\phi^2\right].
\label{eq:infl2}
\eeq

The equations~(\ref{eq:infl1},~\ref{eq:infl2}) are equivalent to the pair
of Hamilton-Jacobi equation~\eqref{eq:jacobi} and 
\beq
\dot\phi=-\dfrac{\mpl^2}{4\pi}H'(\phi).
\label{eq:vel}
\eeq

The Hamilton-Jacobi equation connects the Hubble parameter 
and the value of the potential of the scalar field during the inflation.
In the case when we know the behavior of the Hubble parameter
it is easy to find the potential. The method of flow equations
is entirely based on this fact. 
In contrast, if we know the shape of the potential
and want to reconstruct the behavior of the Hubble parameter, the problem is not as simple.
First of all, as for any first order differential equation,
we would like to have an initial condition $H_0=H(\phi_0)$.
Due to the attractor nature  of
the equation~\eqref{eq:jacobi} its solution does not really
depend on the initial condition $H_0$ (we have found from
numerical simulations that one needs about 6 e-folds to forget the history).
Thus it does not really matter which initial condition we choose. 

Hamilton-Jacobi equation requires that
\beq
H^2(\phi)\ge\frac{8\pi}{3\mpl^2}V(\phi).
\label{eq:ineq}
\eeq
If we are going to use a method such as Runge-Kutta for the integration
of the differential equation\eqref{eq:jacobi}, we might try values of $H$ which would violate
the inequality~\eqref{eq:ineq}.

To avoid this complication, we reparametrize our equation using
a new function $\delta(\phi)$ so that
\beq
H^2(\phi)=\frac{8\pi}{3\mpl^2}V(\phi)\left(1+e^{\delta(\phi)}\right).
\label{eq:defdelta}
\eeq
Then substituting our new definition into equation~\eqref{eq:jacobi}
we get
\beq
H'(\phi)=-\frac{4\pi\sqrt2}{\mpl^2}\sqrt{V(\phi)}e^{\delta(\phi)/2}.
\eeq
Combining this with the expression for $H'$ obtained from the
direct differentiation of $H$ in~\eqref{eq:defdelta}, 
we get a differential equation for $\delta'(\phi)$
\beq
\delta'=-\sqrt{1+e^{-\delta}}\left[\frac{V'}V\sqrt{1+e^{-\delta}}+\frac{4\sqrt{3\pi}}{\mpl}\right].
\label{eq:modif}
\eeq
This equation is much more pleasant to deal with numerically than  
equation~\eqref{eq:jacobi}, since 
it does not have a weird boundary for $\delta$, as $H$ did before.
One can also check the attractor nature of the equation~\eqref{eq:modif},
that it does not remember the prior history.
We see now that in the case when the potential is changing slowly $\delta'\approx0$
and we have
\beq
e^{\delta}=\left[\frac{48\pi}{\mpl^2\left({V'}/V\right)^2}-1\right]^{-1}
\approx\frac{\mpl^2}{48\pi}\left(\frac{V'}{V}\right)^2.
\eeq
We can use this approximate solution of the equation as 
the initial condition for our differential equation since it is
quite close to the true solution and it will make our numerical solution 
evolve into the attractor solution faster.

One can check that the expressions for $\epsilon\sh$, $\eta\sh$ and $^2\xi\sh$
are given by the following formulae
\begin{eqnarray}
\epsilon&=&\frac{\mpl^2}{4\pi}\left(\frac{H'}H\right)^2=\frac3{1+e^{-\delta}},\\
\eta&=&\frac{\mpl^2}{4\pi}\frac{H''}H\notag\\
&=&3+\frac{\mpl}4\sqrt{\frac3\pi}\frac{V'}V
\frac1{\sqrt{e^\delta(1+e^\delta)}},\\
^2\xi&=&\frac{\mpl^4}{16\pi^2}\frac{H'H'''}{H^2}\notag\\
&=&3(\epsilon+\eta)-\eta^2-\frac{3\mpl^2}{8\pi}\frac{V''}V\frac1{1+e^\delta}.
\end{eqnarray}
From these expressions we can expect that in general $\epsilon$ and $\eta$
are continuous functions, whereas $^2\xi$ does not have to be continuous
at points where $V''$ is not continuous.

The condition for inflation to take place ($\ddot a>0$) follows from the derivative
of the Friedman equation
\beq
\frac{\ddot a}a=\frac{8\pi}{3\mpl^2}\left[V(\phi)-\dot\phi^2\right]
=H^2(\phi)(1-\epsilon),
\eeq
or 
\begin{multline}
\frac{\ddot a}a=\frac{8\pi}{3\mpl^2}\left[V(\phi)-\frac{\mpl^4}{16\pi^2}\left(H'(\phi)\right)^2\right]\\
=\frac{8\pi}{3\mpl^2}V(\phi)(1-2e^\delta).
\end{multline}
The first of these two equations implies that the end of inflation happens when
the inequality $\epsilon<1$ is violated.  
The same thing occurs when the inequality $\delta<-\ln2$ is violated in the second equation.
The latter also means that the inflation continues while the kinetic energy of the
inflaton is less than half of its potential energy
\beq
\frac K{\Pi}=\frac{\dot\phi^2/2}{V(\phi)}=e^{\delta(\phi)}<\frac12.
\eeq
Thus we come to a physical definition of our parameter $e^{\delta(\phi)}$
as the ratio of the kinetic energy $\dot\phi^2/2$ to the potential energy $V(\phi)$.

In the next subsection we will be working with inflationary perturbations
and it will not be very convenient for us to work with
the value of the scalar field $\phi$ as an independent variable.
For this purpose we will use the number of e-folds defined as
\beq
\Nt=\ln\frac{(aH)}{(aH)_0}.
\label{eq:Ntilda}
\eeq
Note that this is the actual number of e-folds and
is not the same as $N=\ln (a/a_0)$.
The connection between $\Nt$ and $\phi$ is determined through the derivative
\beq
\frac{d\Nt}{d\phi}=\frac{2\sqrt\pi}{\mpl}\frac{1-\epsilon(\phi)}{\sqrt{\epsilon(\phi)}}.
\eeq

We are almost done describing the background evolution of the universe,
except we have not yet chosen the initial value of the scalar field $\phi_i$. 
We only have the value $\phi_0$ which
corresponds to the moment when the mode $k=0.05$~Mpc$^{-1}$ exits the horizon.
We want to move backwards in time for about 50 e-folds.
Equation~\eqref{eq:modif} has an attractor 
behavior only when we are moving in the positive direction
along the $\phi$-axis. It diverges from the attractor solution 
in the negative direction. 
As a useful trick, let us modify  
equation~\eqref{eq:jacobi} to the following form
\beq
[H'(\phi)]^2=\frac{12\pi}{\mpl^2}\left[\frac{8\pi}{3\mpl^2}V(\phi)-H^2(\phi)\right].
\label{eq:rewritten2}
\eeq
In this form, when we move backwards in time the value of $H(\phi)$ 
is bound by the value of $\sqrt{8\pi V/3\mpl^2}$ from the top and
the solution cannot diverge. 
In addition we temporarily redefine $\delta(\phi)$ to satisfy
\beq
H^2(\phi)=\frac{8\pi}{3\mpl^2}V(\phi)(1-e^{\delta(\phi)}).
\eeq
Thus we get an equation analogous to the equation~\eqref{eq:modif}
\beq
\delta'=\sqrt{1-e^{-\delta}}\left[\frac{V'}V\sqrt{1-e^{-\delta}}+\frac{4\sqrt{3\pi}}{\mpl}\right].
\label{eq:modif2}
\eeq

Equation~\eqref{eq:modif2} does not carry any physical meaning; we just use this equation to
go ``upwards'' to the higher values of the potential, still tracking the
general behavior of $V(\phi)$. If we go backwards in time 50 e-folds using~\eqref{eq:modif2}
and then forward in time 50 e-folds using~\eqref{eq:modif}, we will
not return to the same point $\phi_0$, since the behavior of $\delta(\phi)$ in 
the equation~\eqref{eq:modif2} is determined by the area which is to the right
of the current value of $\phi$ and in the equation~\eqref{eq:modif}
is determined by the area which is on the left side.
Nevertheless, this method gives us a good estimate of what initial
value of $\phi_i$ we should take.

It is also worth mentioning that this approach is not more \
difficult to deal with 
than the inflationary flow equations (\ref{eq:flow1}--\ref{eq:flowlast}).

\subsection{Perturbation equations}
\subsubsection{Scalar mode}
The algorithm for finding the scalar mode primordial power spectrum
is described in the main text (see equation~\eqref{eq:modeeq} and below).
Here we will just mention some technical details.

Equation~\eqref{eq:modeeq} is not very convenient to solve in its current form.
First of all we would like to set the independent variable, the conformal time $\tau$,
in such a way that $\tau\to-0$ as inflation goes on.
In this case we would be able to numerically integrate 
equation~\eqref{eq:modeeq} up to as small
values of $\tau$ as we want. But
in numerical realizations we cannot really choose such an initial value of $\tau_i$
that gives us $\tau\to-0$ at the end of the inflation.
Suppose that at the end of the inflation we have $\tau\to1-0$. 
In this case the numerical error on $\tau$ will be of the order of $10^{-15}$
which is a reasonable machine precision. Hence the limit
on corresponding $d\tau$ is of the same order and we can explore
the range of changing the scale factor $a$ from $\sim1$ to $\sim10^{15}$,
i.e. about 35 e-folds. This might be enough, but to be safe
we will use a different independent variable, the true number of 
e-folds $\Nt$ defined by equation~\eqref{eq:Ntilda} which is the same as
\beq
d\Nt=\frac{d(aH)}{aH}.
\eeq

Then the mode equation~\eqref{eq:modeeq} can be rewritten as
\begin{multline}
(1-a)\frac{d^2u_k}{d\Nt^2}+(1+b)\frac{du_k}{d\Nt}\\
+\left[\left(\frac k{k_0}\right)^2e^{-2(\Nt-\Nt_0)}-2(1+c)\right]u_k=0,
\label{eq:modeeq_modif}
\end{multline}
where coefficient $a$, $b$ and $c$ can be exactly expressed
through $\epsilon\sh$, $\eta\sh$ and $^2\xi\sh$ as
\beqa
\label{eq:coeffsa}
a&=&2\epsilon-\epsilon^2,\\
b&=&-2\epsilon-\epsilon^2+2\epsilon\eta,\\
c&=&\epsilon-\frac32\eta+\epsilon^2-2\epsilon\eta+\frac12\eta^2+\frac12~^2\xi.
\label{eq:coeffsc}
\eeqa
In equation~\eqref{eq:modeeq_modif}, $k$ is the wavelength of the interest,
while $k_0$ and $\Nt_0$ are constants conveniently chosen
for normalization purposes.

Further, equation~\eqref{eq:modeeq} has a solution
\beq
u_k\propto \frac1{\sqrt{2k}}e^{-ik\tau}
\eeq
at the beginning of the inflation when $\tau\to-\infty$ and $k^2\gg \dfrac1z\dfrac{d^2z}{d\tau^2}$.
We also know the approximate behavior of $u_k$ at later times when $\tau\to-0$ 
and $k^2\ll\dfrac1z\dfrac{d^2z}{d\tau^2}$:
\beq
u_k\propto z.
\eeq

Thus it is natural to decompose $u_k$ into growing and oscillating parts
\beq
u_k=e^{A+i\phi},
\eeq
where both functions $A$ and $\phi$ are real functions of conformal time $\tau$
or of the true number of e-folds $\Nt$. Then 
the equation~\eqref{eq:modeeq_modif} can be split into
4 ordinary differential equations with 2 new functions $A_p$ and $\phi_p$ defined as below
\beqa
\frac{dA}{d\Nt}&=&A_p,\label{eq:eq1}\\
\frac{dA_p}{d\Nt}&=&
\frac{-\left.({k^2}/{k_0^2})\right.e^{-2\Nt}+2(1+c)-(1+b)A_p}{1-a}\notag\\
&&-(A_p^2-\phi_p^2),
\label{eq:eq2}\\
\frac{d\phi}{d\Nt}&=&\phi_p, \label{eq:eq3}\\
\frac{d\phi_p}{d\Nt}&=&
-\phi_p\frac{(1+b)+2(1-a)A_p}{1-a}. \label{eq:eq4}
\eeqa

This system of differential equations looks a bit more complicated than the single
equation~\eqref{eq:modeeq}, but it is actually much easier to solve numerically.
Indeed, at earlier times we have $dA/d\Nt\equiv A_p=0$. This instaneously
gives us the initial condition on $d\phi/d\Nt\equiv \phi_p$ from~\eqref{eq:eq2}
\beq
\phi_p^2=\frac{\left.({k^2}/{k_0^2})\right.e^{-2\Nt}-2(1+c)}{1-a}
\eeq
as $\tau\to-\infty$, i.e. $\Nt\to-\infty$. To be consistent with
the initial condition on $$u_k\propto\dfrac1{\sqrt{2k}}e^{-ik\tau}$$ as $\tau\to-\infty$
we also require that $$A=-\frac12\ln k$$ as $\Nt\to-\infty$. As the inflation continues,
the terms 
$$-\dfrac{k^2}{k_0^2}\dfrac{e^{-2\Nt}}{1-a}$$ 
and $\phi_p^2$
will balance each other on the right hand side of the equation~\eqref{eq:eq2} 
until $A_p$ is not negligible in comparison to $1$ in equation~\eqref{eq:eq4}.
Thus, around $\Nt=\ln(k/k_0)$ the oscillating part $\phi_p$
will decrease more rapidly than before, finally exponentially dropping to zero. 
At the same time $A_p$, and therefore $A$, start exponentially growing.
The final power spectrum is given by
\beq
\mathcal P_k=\frac{k^3}{2\pi^2}\left|\frac{u_k}z\right|^2
\propto \frac{k^3}{2\pi^2} e^{2A_k}.
\eeq
Thus we even do not need information about the phase $\phi$ and we can freely
drop equation~\eqref{eq:eq3} from our system. Also while being
in the stage of inflation where $u_k$ has an oscillatory behavior,
if one were to use the usual method without our substitution, one would have
to find the values of $u_k$ for at least 6 points per oscillation period.
However with our substitution, we easily pass this area, which
does not have any interest for us since we analytically know the behavior of $u_k$ here,
and therefore move directly to the place where we cannot solve it analytically. 
By our estimates this technique gives a gain of a factor of 10 in computational time,
which is of particular interest if one wants to calculate the power spectrum
for e.g. 100 wavemodes.

\subsubsection{Tensor mode}
The calculation of the tensor mode power spectrum of perturbations
is absolutely analogous to the one for scalars, except instead of 
equation~\eqref{eq:modeeq} one has to solve 
\beq
\frac{d^2u_k}{d\tau^2}+\left(k^2-\frac1a\frac{d^2a}{d\tau^2}\right)u_k=0
\eeq
with the same initial condition $$u_k(\tau)\to\dfrac1{\sqrt{2k}}e^{-ik\tau}$$ as $\tau\to-\infty$,
where $a$ is the usual scale factor of the Friedman universe. One can show that 
\beq
\frac1a\frac{d^2a}{d\tau^2}=2a^2H^2\left(1-\frac12\epsilon\right).
\eeq

Mode equations for the amplitude and the phase of the wave (\ref{eq:eq1}-\ref{eq:eq4}) 
of the tensor mode look similar
except in the equations (\ref{eq:coeffsa}-\ref{eq:coeffsc}) where we
have to change $c$ to $d$ defined as
\beq
d=-\frac12\epsilon.
\eeq

\end{document}